\documentclass[aps,prl,reprint]{revtex4-2}

\usepackage{graphicx} 
\usepackage{xcolor}
\usepackage{amsmath}
\usepackage[colorlinks,citecolor=blue,urlcolor=blue,bookmarks=false,hypertexnames=true]{hyperref} 

\begin{document}

\title{Mechanochemical Morphodynamics of Active Bacterial Cells}

\author{Joydip Chaudhuri}
\email[]{joy@che.iitkgp.ac.in}
\affiliation{Department of Chemical Engineering, Indian Institute of Technology Kharagpur, Kharagpur 721302, West Bengal, India}



\begin{abstract}
Bacterial cells exhibit a diverse array of shapes and sizes, largely governed by their cell walls in conjunction with cytoskeletal proteins and internal turgor pressure. The present study develops a theoretical framework for modeling the shape dynamics of actively expanding bacterial cell walls, grounded in the concept of minimal energy dissipation. In the context of a bacterial cell wall, dissipative forces are generated through the insertion of peptidoglycan (PG) strands, while driving forces stem from alterations in mechanochemical energy, crucial for sustaining the cell wall's shape. The interplay between mechanical and chemical energies facilitates in evaluating the free energy landscape and helps in predicting the homeostasis of the bacterial cell size. The size limit derived through linear stability analysis (LSA) of a model system accurately mirrors the phase diagram produced by the theoretical model. Nonetheless, given the cell wall's intricate molecular architecture, a more detailed constitutive model is expected to provide more precise quantitative insights.
\end{abstract}

\maketitle


Active materials, inherently existing outside of thermal equilibrium, possess the ability to transform stored internal or external free energy into organized motion or the exertion of mechanical forces. This phenomenon is evident in the biological realm, particularly in bacterial colonies, cell membranes, and the cytoskeleton of biological cells, which all serve as examples of such active matters. The growth dynamics of these living active matters is influenced by various sub-cellular biochemical processes which ultimately determine a mechanochemically stable shape and structure. Understanding the physics associated with the growth dynamics of these living active systems based on the molecular processes and environmental factors may shed light on many biological processes, including healing of wounds, tissue morphogenesis, tumor metastasis, and development of plant and bacterial cell walls.  In this context, the growth of cells and bacterial cell walls, driven by biochemical assembly reactions, has been extensively studied \cite{vollmer2008murein,hett2008bacterial,pasquina2020architecture, Mugler2024}.

Based on the pioneering work of Thompson which deals with the forms in nature from a perspective of physical arguments \cite{thompson19171942}, Koch developed the so-called surface stress theory \cite{koch1983surface,koch1994problem}, which suggests that the bacterial cell shapes are predominantly determined by the surface stresses in the cell wall \cite{koch1983surface,koch1994problem}. Recently, a number of theoretical models have been proposed to understand shape and growth of bacterial cell walls \cite{boudaoud2003growth,jiang2010morphology,amir2012dislocation,banerjee2016shape} which focuses on some critical factors such as, growth by plastic deformations as surface stresses exceed a critical value \cite{boudaoud2003growth}, elastic growth of PG networks driven by assembly reactions \cite{jiang2010morphology}, dislocation driven growth of partially ordered PG structures \cite{amir2012dislocation}, and exponential longitudinal growth, stability, and division kinetics in rod-like cells \cite{banerjee2016shape}. These models effectively forecast the rate of cell growth, but they fall short in calculating the potential limit to cell size before division, as they consider only a portion of the mechanical forces at play, not the full spectrum. The present study addresses this gap by introducing a comprehensive theoretical framework for modeling the shape dynamics of active cell walls. A phase diagram obtained from this framework accurately predicts the steady-state size of a cell and corresponds with the cell size limits found in experimental data, taking into account the full range of mechanochemical forces in the cell wall.


We express the cellular wall's geometry through a number of degrees of freedom represented by generalized shape parameters $m_i$ ($i=1,2,3,...$) and their corresponding growth velocities $d{m_i}/dt$ or hereafter denoted as $d_{t}{m_i}$. For instance, a sphere's shape parameter involves its radius, while a cylinder requires two degrees of freedom, namely its radius and length. The total free energy of the cell wall is a function of these generalized shape parameters, denoted as $G({m_i})$. Changes in the cell wall's geometry are propelled by generalized driving forces, $F_i$, which are the derivatives of the free energy ($G$) function with respect to $m_i$, 
\begin{equation}\label{1}
    F_i=-\partial_{m_i}{G(m_i)}, \forall{i}.
\end{equation}
The equilibrium configuration of the cell wall is established by minimizing the free energy, expressed as $\displaystyle {\partial_{m_i}{G(m_i)}}=0, \forall{i}$, indicating that ${F_i} = 0$. It should be noted here that, this is the simplest conceivable model to describe such a non-equilibrium phenomenon which follows the linear Onsager theory. However, it does not faithfully represents the actual dynamics of the system, as it does not include non-linear effects and cross-correlations. In such a scenario where the dynamics follows linear response, it is expected that the growth velocity ($d_{t}{m_i}$) or the rate of growth of the shape parameters ($m_i$) is proportional to the driving force ($F_i$). Consequently, the growth velocity associated with the shape parameter $m_i$ can be represented as follows:
\begin{equation}\label{2}
    {d_{t}m_i}\propto {F_i} \Rightarrow {d_{t}m_i} = -Q_i\partial_{m_i}{G(m_i)}, \forall{i},
\end{equation}
where, $Q_i$ is the phenomenological growth constant corresponding to the shape parameter $m_i$. Equation (\ref{2}) embodies the well-known constitutive law of `Newtonian' flow, where the internal stress (on the right-hand side) is proportional to the rate of strain (on the left-hand side). In the context of that analogy, the constant $Q_i$ is akin to `Newtonian' viscosity of the cell. Equation (\ref{2}) can also be mathematically mapped to the overdamped motion of a particle in a potential. The equations of motion of the shape parameters (\(m_i\)) have also been analyzed in order to evaluate the rate of growth characteristics, although the free energy landscape described above is sufficient in order to predict the potential cell size limit. 

The overall change in free energy ($G$) can be expressed as,
\begin{equation}\label{4}
dG=dE-\epsilon{dA},
\end{equation}
where \(dE\) represents the alteration in the mechanical energy of the cell wall network and the active chemical energy liberated during the incremental growth of the cell wall area ($dA$) is $\epsilon{dA}$. The parameter, $\epsilon$ ($>0$) signifies the effective chemical potential for cell surface growth. The term $dE$ is contingent upon the shape and size of the cell wall, suggesting that there might exist a specific size and shape for the cell where the augmented mechanical strain energy precisely counterbalances the active chemical energy, resulting in $dG=0$. This equilibrium implies a $mechanochemical$ energy balance in the system. Once this configuration is attained, assembly and disassembly reactions harmoniously equate, causing the cessation of cell wall growth, thus reaching a potential maximum cell size before its eventual division. Conventionally, this approach of achieving a size homeostasis is termed as ``sizer'' mode of cell size regulation \cite{amir2014,Mugler2024}. In sizer regulation, the cell division cycle does not proceed until a minimum \textit{target} size is reached, irrespective of its initial birth size or the time required to grow, requiring the cell to monitor its own size by active cell size control strategies. In the present minimal model, these active controls including the insertion of PG strands in the cell wall are all bundled together as chemical potential ($\epsilon$) in equation (\ref{4}).

The mechanical energy ($E$) related to the growing cell wall in equation (\ref{4}) is the collective outcome of various components. These include all possible mechanical energies associated with the cell wall such as, the influence of an internal turgor pressure ($P_{tp}$), which works to expand the cell volume ($V$), the resistance posed by surface tension ($\sigma$) against an augmentation in the cell surface area, and the mechanical energies associated with cell wall ($E^{cw}$) itself, governing the cell's shape. In other words,
\begin{equation}\label{5}
E = -\int{P_{tp}dV}+\int{\sigma d{A}}+{E^{cw}}.
\end{equation}
The surface tension ($\sigma$) is determined by the stored elastic energy per unit area of the cell wall, potentially mitigated by favorable PG interactions at the surface. Components of $E^{cw}$ encompass MreB bundles that regulate cell width \cite{jones2001control,figge2004mreb}, FtsZ filaments responsible for propelling cell wall constriction \cite{erickson1996bacterial}, and crescentin bundles that govern curvature in \textit{C. crescentus} cells \cite{ausmees2003bacterial}. In this study, $E^{cw}$ is articulated as the amalgamation of various stretching and bending energies, in contrast to earlier theoretical models that considered either only stretching or bending energies \cite{ausmees2003bacterial,banerjee2016shape,jiang2010morphology}. As we aim to generalize the theory regarding the biophysics of the bacterial cell wall, we have limited our consideration to simpler cellular geometries, such as spherical and cylindrical, for the sake of simplicity in our proposed model.

For simple spherical shape with small deformations, the cell wall mechanical energy ($E^{cw}_{S}$) can be expressed as,
\begin{equation}\label{6}
E^{cw}_{S}=4\pi P_SR_S^4,
\end{equation} 
where, $\displaystyle P_S = \frac{P_{tp}^2}{8h\left(\lambda+\mu\right)}$ and $R_S$ is the radius of the spherical cell, serving as the sole shape parameter for a spherical geometry \cite{jiang2010morphology}. The Lame coefficients, \(\lambda\) and \(\mu\) are related to the Young's modulus (\(Y\)) and the Poisson ratio (\(\nu\)) of the cell wall according to the following expressions, $\lambda={\nu{Y}}/{\left(1-\nu^2\right)}, \hspace{2pt}\mathrm{and}\hspace{2pt} \mu={Y}/{2\left(1+\nu\right)}$ \cite{jiang2010morphology}. For growing spherical bacterial cells like, \textit{S. pneumoniae} or \textit{S. aureus}, where cytoskeletal bundles such as MreB are known to be absent, the effective free energy due to the bending rigidity can be neglected in calculating the total free energy. For such cells, the calculation of total free energy of the cell wall from equations (\ref{4}) $-$ (\ref{6}) is simple as follows,
\begin{equation}\label{7}
    G_S = 4\pi \left[P_S{R_S^4}-\left(\epsilon-\sigma\right){R_S^2}-\frac{1}{3}P_{tp}R_S^3\right].
\end{equation}
Therefore, the radial growth velocity of the spherical cell can be evaluated using equation (\ref{2}) as,
\begin{equation}\label{8}
    {d_{t}R_S}=4\pi{Q_R^S}R_S\biggl[-4P_SR_S^2+2\left(\epsilon-\sigma\right)+R_S\frac{d\sigma}{dR_S}+P_{tp}R_S\biggr],
\end{equation}
where, $Q_R^S$ represents the corresponding proportionality constant.

If the cell wall behaves akin to a plastic material, where $\sigma$ is a constant \cite{boudaoud2003growth}, a critical radius, referred to as the steady-state radius ($R_S^s$), can be determined by setting $\displaystyle {d_{t}R_S} = 0$. This condition implies that,
\begin{equation}\label{9}
    R_S^s=\frac{P_{tp}}{8P_S}\left[1+\sqrt{1+\frac{32P_S\left(\epsilon-\sigma\right)}{P_{tp}^2}}\right].
\end{equation}
Considering, the active energy ($\epsilon$) to be an effective active growth pressure and combining it with the turgor pressure ($P_{tp}$), equation (\ref{9}) asymptotically reduces to, $R_S^s=2\sigma/P_{tp}$, which is analogous to Laplace's law \cite{gennes2004capillarity}. Furthermore, it may be noted here that, equation (\ref{9}) asymptotically reduces to the steady state radius predicted by the earlier phenomenological theory \cite{jiang2010morphology}, if we neglect the effect of surface tension and pressure-volume work in equation (\ref{7}). 

However, if the surface tension arises from the elastic strain energy stored in the pressurized spherical shell, the expression for $\sigma$ becomes: $\displaystyle \sigma = P_{tp}^2R_S^2/8\sigma_0$, with $\sigma_0 = Y h/2(1-\nu)$ \cite{bower2009applied}. In this case, where $\sigma = f(R_S)$, the steady state radius ($R_S^s$) can be obtained from equation (\ref{8}) as,
\begin{equation}\label{10}
R_S^s = \frac{1}{2P_E}\left[1+\sqrt{1+\frac{8\epsilon{P_E}}{P_{tp}}}\right], 
\end{equation}
where, $\displaystyle P_E = \left[\left({4P_S}/{P_{tp}}\right)+\left({P_{tp}}/{2\sigma_0}\right)\right]$. Under these circumstances too, the cell radius converges asymptotically to the critical value $2\sigma_0/P_{tp}$, analogous to Laplace's law \cite{gennes2004capillarity}. This situation is particularly relevant to spherical bacteria that maintain a consistent cell size before entering the cell division phase \cite{kuru2012situ}. Consequently, a plastic cell wall can sustain continuous growth given optimal nutrient availability and division inhibition. Conversely, an elastic cell wall cannot sustain growth beyond a specified threshold size, $R_S^s$, as determined by equation (\ref{10}), at which point it promptly enters the cell division phase following the ``sizer" regulation. Equations (\ref{9}) and (\ref{10}) also suggest that a plastic cell wall can withstand the internal turgor pressure of a larger cell volume than an elastic cell wall \cite{zhang2021molecular,fratzl2022mechanical}.

In case of simple cylindrical cells such as, \textit{E. coli}, the mechanical energy of the cell wall ($E^{cw}_{C}$) encompasses both stretching and bending free energy components to uphold the cylindrical cell structure. For cylindrical shapes characterized by small deformations, the mechanical energy of the cell wall can be expressed in a simplified form, as indicated in:
\begin{equation}\label{11}
E^{cw}_{C} = 2\pi {L_C} {R_C}\left[P_CR_C^2 - \frac{k}{2}\left(\frac{1}{R_C}-\frac{1}{R_0}\right)^2\right],
\end{equation}
where, $\displaystyle P_C=\frac{P_{tp}^2\left(\lambda+10\mu\right)}{32h\mu\left(\lambda+\mu \right)}$, $k$ is the circumferential bending rigidity of the cell wall, $R_0$ is the preferred radius of cross section of cell wall. Furthermore, $R_C$ is the radius of the cylindrical cell and $L_C$ is the length of the cylinder region, which are the two shape parameters associated with a cylindrical geometry.
For such rodlike cells, the total free energy from equations (\ref{4}), (\ref{5}) and (\ref{11}) can be expressed as,
\begin{equation}\label{12}
\begin{split}
    G_C &= 2\pi L_C R_C \biggl[ P_C R_C^2 - (\epsilon - \sigma) - \frac{1}{2} P_{tp} R_C \\
    & \quad - \frac{k}{2} \left( \frac{1}{R_C} - \frac{1}{R_0} \right)^2 \biggr]
\end{split}
\end{equation}

The cell poles being rigid and inert are neglected in this calculation \cite{boudaoud2003growth,ausmees2003bacterial}. Considering $\sigma$ to be constant in this case (plastic cell walls), the growth equations in the radial and axial directions can therefore be expressed using equation (\ref{2}) as:

\begin{equation}\label{13}
\begin{split}
    d_{t}R_C &= -2\pi Q_R^C L_C \biggl[ 3P_C R_C^2 - (\epsilon - \sigma) - P_{tp} R_C \\
    & \quad - \frac{k}{2} \left( \frac{1}{R_C} - \frac{1}{R_0} \right)^2 + \frac{k}{R_C} \left( \frac{1}{R_C} - \frac{1}{R_0} \right) \biggr]
\end{split}
\end{equation}

and,

\begin{equation}\label{14}
\begin{split}
    d_{t}L_C &= -2\pi Q_L^C \biggl[ P_C R_C^3 - (\epsilon - \sigma) R_C - \frac{1}{2} P_{tp} R_C^2 \\
    & \quad - \frac{k}{2} R_C \left( \frac{1}{R_C} - \frac{1}{R_0} \right)^2 \biggr]
\end{split}
\end{equation}

It is noteworthy that, $Q_R^C$ and $Q_L^C$ in equations (\ref{13}) and (\ref{14}), represent the proportionality constants corresponding to the two shape parameters, $R_C$ and $L_C$, respectively. The radius at the steady state ($R_C^s$) can be obtained by considering $\displaystyle {d_{t}R_C} = 0$. This implies that, $R_C^s$ is one of the positive real roots of the following fourth order polynomial of $R_C$, 
\begin{eqnarray}\label{15}
- \frac{k}{2}\left(\frac{1}{R_C}-\frac{1}{R_0}\right)^2 + \frac{k}{R_C}\left(\frac{1}{R_C}-\frac{1}{R_0}\right) \nonumber \\
+3P_CR_C^2-P_{tp}R_C-\left(\epsilon-\sigma\right) 
 &=& 0.
\end{eqnarray}
The temporal equations (\ref{8}) and (\ref{13}) depict the dynamics governing the evolution of the radius for spherical and cylindrical cells, respectively. This finding closely corresponds to predictions made in earlier theories, as illustrated in Figures \textcolor{red}{S1} and \textcolor{red}{S2} of the supplementary information (\textcolor{red}{SI}) \cite{jiang2010morphology,banerjee2016shape}. According to equation (\ref{14}), the temporal evolution of cell length ($L_C$) is expected to be linear with respect to time ($t$), as the right-hand side of equation (\ref{14}) remains independent of $L_C$. This observation also aligns closely with predictions from previous theories, as shown in Figure \textcolor{red}{S3} of \textcolor{red}{SI} \cite{jiang2010morphology,banerjee2016shape}. However, the length and time scales of the dynamics of these shape parameters from the present study differ significantly from those in previous literature, yet align closely with experimental findings \cite{thwaites1991mechanical,takeuchi2005controlling,cabeen2009bacterial}.

\begin{figure}
\centering
\includegraphics[width=1.0\linewidth]{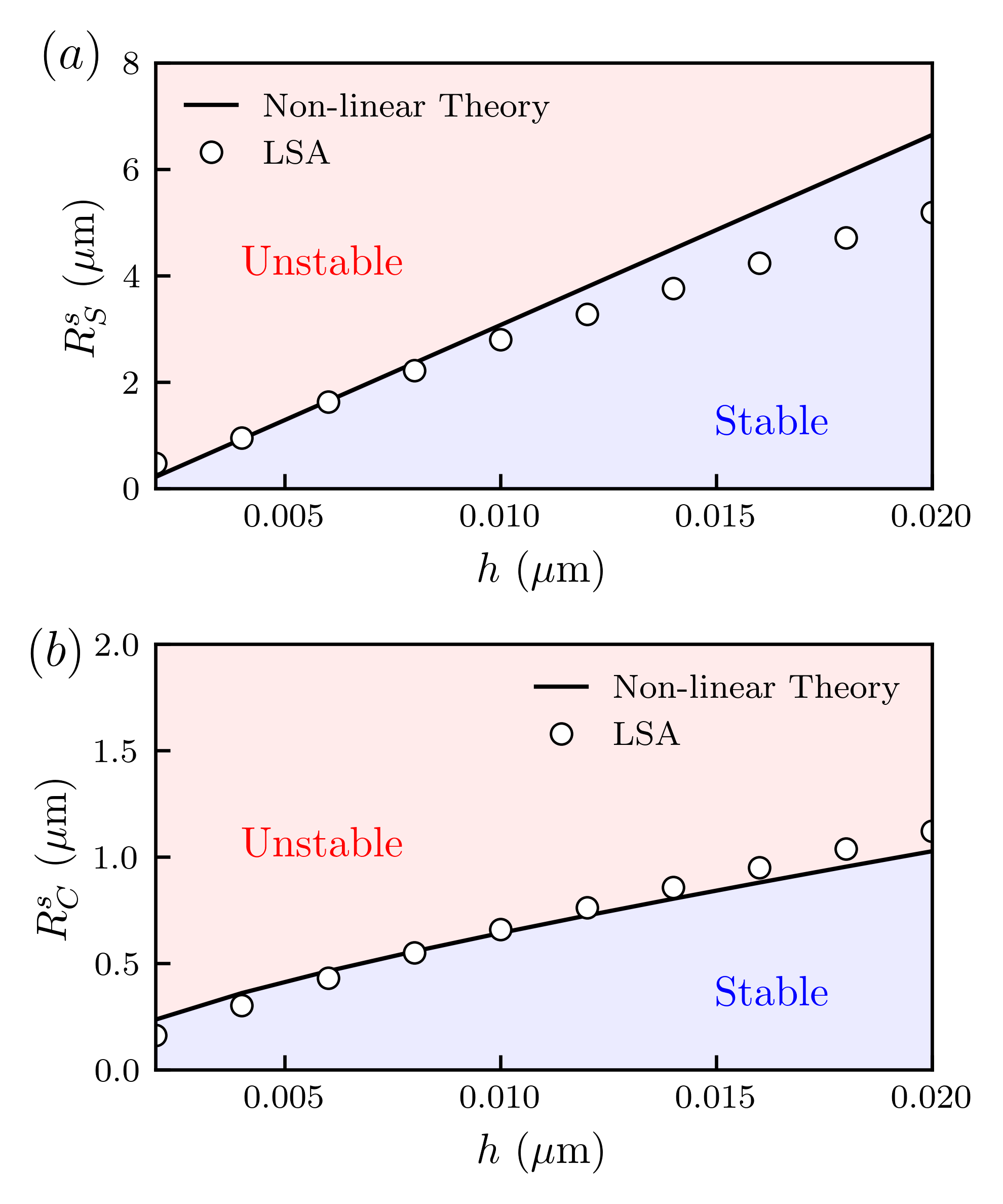}%
\caption{\label{fig1}Plot shows the variation of the steady state radius of ($a$) spherical ($R_S^s$) and ($b$) cylindrical cells ($R_C^s$), respectively, for different cell wall thicknesses ($h$) from the non-linear theory and LSA.}
\end{figure}

\begin{figure}
\centering
\includegraphics[width=1.0\linewidth]{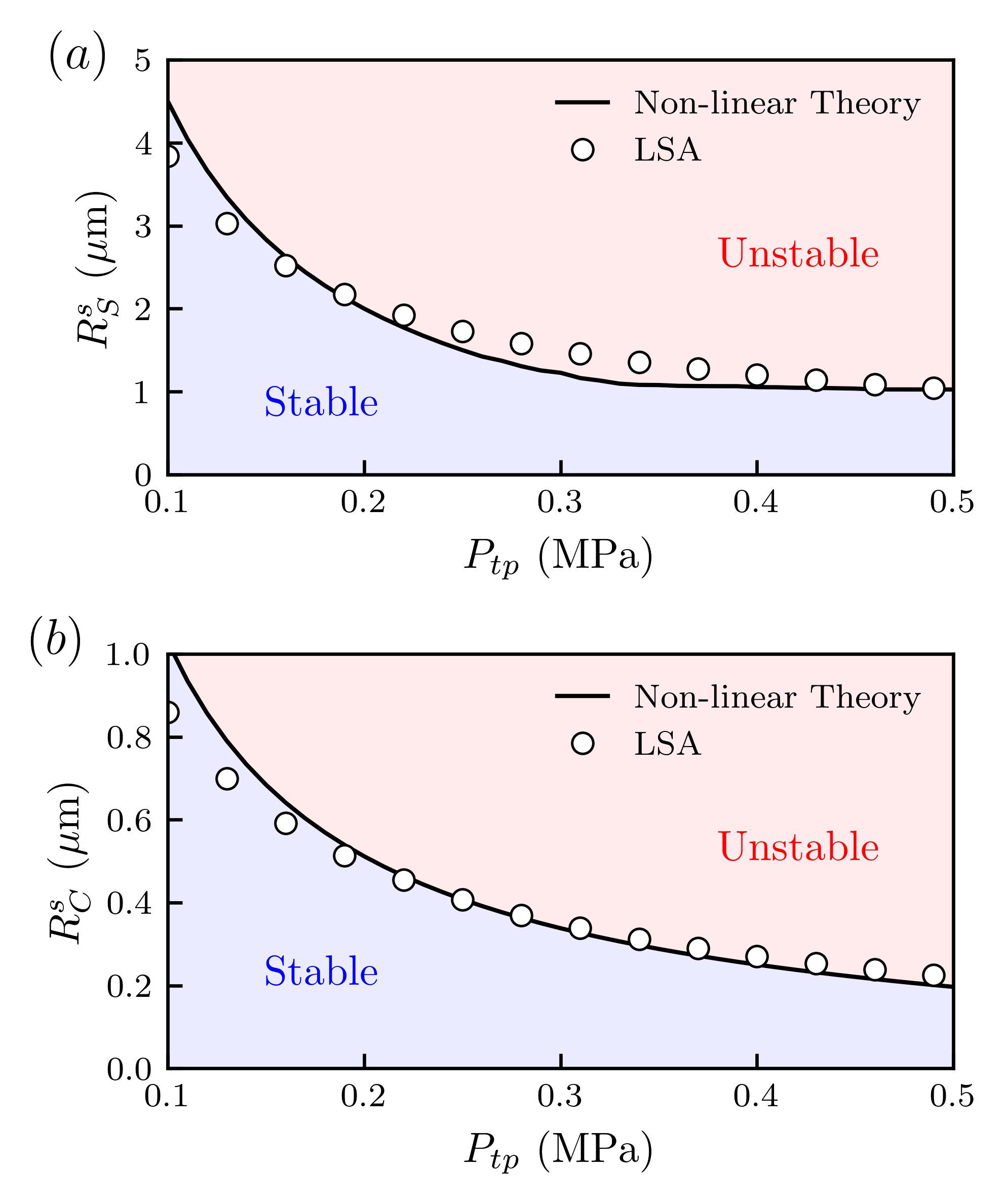}%
\caption{\label{fig2}Plot shows the variation of the steady state radius of ($a$) spherical ($R_S^s$) and ($b$) cylindrical cells ($R_C^s$), respectively, for different turgor pressures ($P_{tp}$) from the non-linear theory and LSA.}
\end{figure}

Having examined the growth and shape dynamics under stable environmental conditions, we now explore how spherical and cylindrical cells adjust their growth dynamics in reaction to morphological disturbances. Experimental observations reveal that the introduction of A22, a reversible inhibitor of the bacterial cell wall protein MreB which leads to the disassembly of the same, results in the formation of wave-like bulges on the cell wall \cite{jiang2011mechanical}. We conduct LSA on the steady-state equations (\ref{6}) to (\ref{14}), introducing perturbations in both the steady-state cell radii ($R_S$ and $R_C$) along the axial direction $z$ for both spherical and cylindrical cells. The perturbations take the form of following normal linear modes:

\begin{align}
R_S &= {R}_S^s (z,t) + \widetilde{R}_S(t) e^{\omega{t}+ikz}, \\
R_C &= {R}_C^s (z,t) + \widetilde{R}_C(t)e^{\omega{t}+ikz},
\end{align}
where, $\omega$ is the complex growth rate, $k$ $(> 0)$ is the wave number, and $\widetilde{R}_S(t) \ll {R}_S^s (z,t)$, $\widetilde{R}_C(t) \ll {R}_C^s (z,t)$ are the amplitudes of the perturbations and $i = \sqrt{-1}$. A perturbation is unstable when $\omega > 0$ such that wave-like bulges nucleate on the cell-wall with growing amplitude, whereas, it is stable when $\omega < 0$ such that any wave-like bulges will essentially die down with time and the cell will retain its shape. Neutral stability curves are evaluated from $\omega = 0$, which concurrently yields a set of steady-state radii ($R^s$) evaluated through LSA (details in \textcolor{red}{SI}). We employed LSA to assess $R^s$ for both spherical ($R_S^s$) and cylindrical ($R_C^s$) bacterial cells across various crucial thermodynamic parameters, as depicted in Figures \textcolor{red}{S4} $-$ \textcolor{red}{S6} of \textcolor{red}{SI}. 

\begin{figure}
\centering
\includegraphics[width=1.0\linewidth]{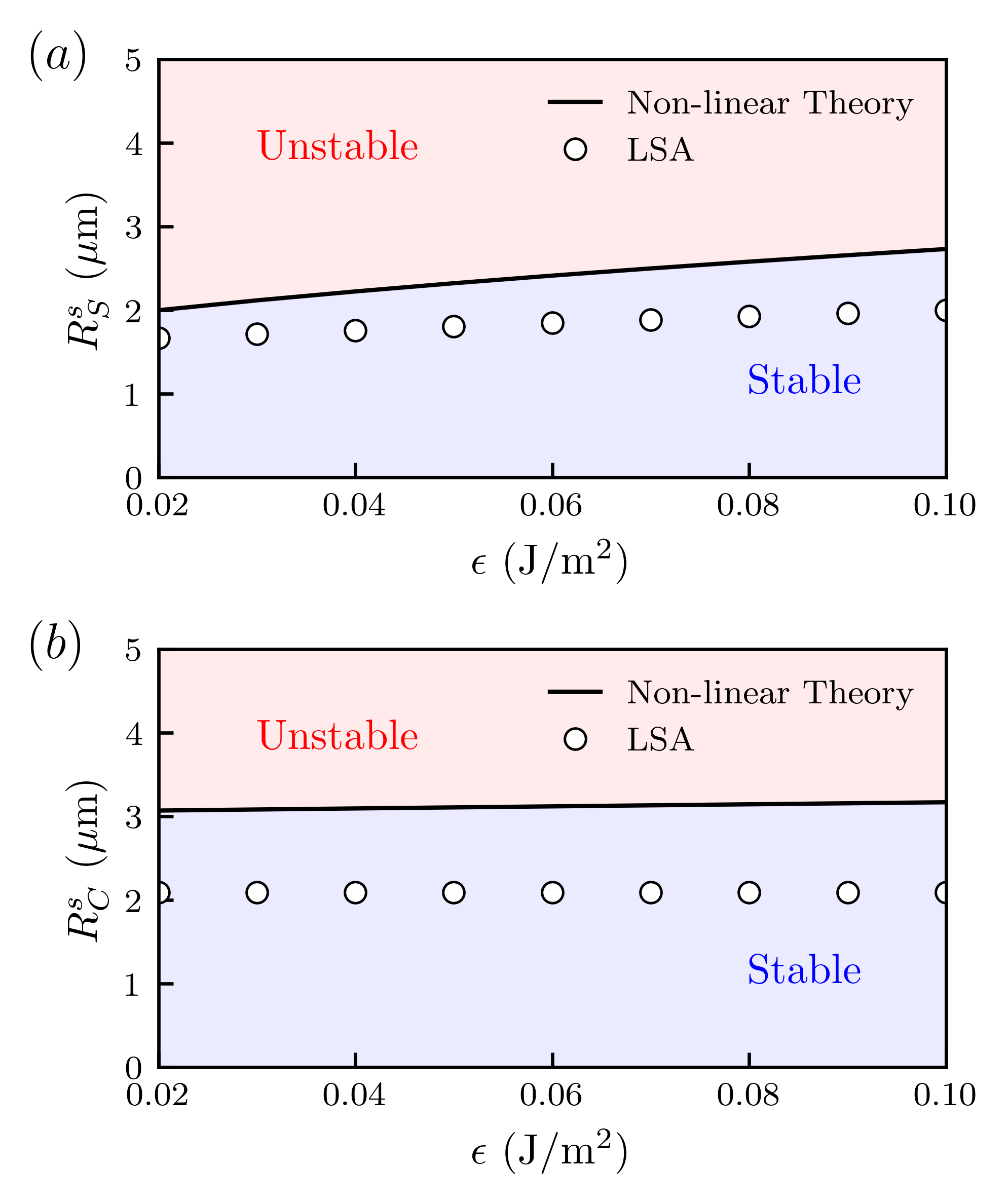}%
\caption{\label{fig3}Plot shows the variation of the steady state radius of ($a$) spherical ($R_S^s$) and ($b$) cylindrical cells ($R_C^s$), respectively, for different energy released ($\epsilon$) from the non-linear theory and LSA.}
\end{figure}

Additionally, we compared the values of $R^s$ obtained from LSA with those derived from the non-linear theory, as detailed in equations (\ref{1}) to (\ref{15}). Figures \ref{fig1} to \ref{fig3} present a phase diagram illustrating the variation of $R^s$ for spherical and cylindrical cells across different cell wall thicknesses ($h$), turgor pressures ($P_{tp}$), and active energy released ($\epsilon$). These figures depict the maximum achievable cell size or the \textit{target} size for ``sizer" regulation based on various parameters ($h, P_{tp}$ and $\epsilon$) predicted by both full non-linear theory and LSA. Consequently, they contribute to the development of a phase diagram that outlines regions of instability and stability concerning cell size. Notably, the LSA exhibits commendable accuracy in predicting the variation of $R^s$ when compared to the outcomes derived from the full non-linear theory.

A key characteristic of the growth of rod-like cylindrical bacteria is their ability to elongate while maintaining a constant radius, implying a complex relationship between growth energy and their shape. The present research mirrors this complexity, as shown in Figures \ref{fig3}($b$) and \textcolor{red}{S6}($b$) of \textcolor{red}{SI}. Figure \ref{fig3}($b$) illustrates that upon reaching the steady-state radius (\(R_C^s\)), the cylindrical cell's radius remains nearly constant, even with an increased release of active energy (\(\epsilon\)) inside the cell. This implicitly suggests that cylindrical cells with a steady-state radius (\(R_C^s\)) experience only longitudinal growth, maintaining the cell's radius.

Furthermore, the shape and size of cells obtained from LSA as well as from non-linear theory closely align with the cell sizes observed in experimental studies \cite{thwaites1991mechanical,young2006selective,margolin2009sculpting,tilby1978detergent,woldringh1977morphological}. For example, typical values of parameters for typical bacterial cells are: $P_{tp} = 0.3$ MPa (for Gram-negative) $-$ $1.5$ MPa (for Gram-positive), $Y = 20-70$ MPa, $h = 3-20$ nm, $\nu =0.3$ (assumed) \cite{jiang2010morphology}, $\epsilon = 0.05$ J/m$^2$ (for Gram-negative) $-$ $0.25$ J/m$^2$ (for Gram-positive), $Q_i = 0.01$ m$^2$/J s \cite{jiang2010morphology}, $k = 0.03$ MPa$\mu$m$^3$, $R_0 = 0.38 - 0.43$ $\mu$m, and $\sigma = 20\times 10^{-3}$ $-$ $120\times 10^{-3}$ N/m \cite{allard2009force,thwaites1991mechanical,young2006selective}. With these parameters, $\langle R_C^s \rangle$ (mean value of $R_C^s$) is calculated to be $\approx 0.65$ $\mu$m (LSA: $\approx 0.66$ $\mu$m, non-linear theory: $\approx 0.64$ $\mu$m, as per Fig. \ref{fig1}) for \textit{E. coli} cells with cell wall thickness $h =10$ nm, which is in good agreement with the experimental value of $\approx 0.38 - 0.72$ $\mu$m \cite{thwaites1991mechanical,young2006selective,young2010bacterial,margolin2009sculpting,tilby1978detergent,woldringh1977morphological}. As another illustration, $\langle R_S^s \rangle$ (mean value of $R_S^s$) is evaluated for a typical \textit{cocci} bacterial cell, which is found to be $\approx 1.025$ $\mu$m (LSA: $\approx 1.03$ $\mu$m, non-linear theory: $\approx 1.02$ $\mu$m, as per Fig. \ref{fig2}) for $P_{tp} = 0.5$ MPa, aligning well with experimental value of $\approx 0.25 - 1.25$ $\mu$m \cite{young2006selective,young2010bacterial}. These resemblances indicate that the phase diagram in Figures \ref{fig1} to \ref{fig3}, as anticipated by both LSA and non-linear theory, can offer reliable preliminary approximation of size homeostasis of a bacterial cell prior to division.

In summary, we present a theoretical model for analyzing the shape dynamics and potential size limits of actively growing cell walls, based on minimal energy dissipation and mechanochemical energy balance. Active forces originate from proteins propelling cell wall growth, while mechanical forces arise from tensions in the PG cell wall and the associated cytoskeletal bundles. The cell size limit determined through LSA consistently predicts the phase diagram derived from the non-linear theory. Additionally, the calculated length and time scales of such active cells align closely with experimental findings for basic cell shapes. The dynamic and ever-evolving nature of bacterial cell walls, with their fluctuating mechanical properties due to molecular defects and random forces, must be acknowledged \cite{Mugler2024}. Our model is primarily effective for timescales of observable cell wall growth kinetics and not effective for stochastic and molecular timescales. Future efforts will focus on integrating non-linear, stochastic and spatiotemporal variations to deepen our understanding of such active morphodynamics of different cells.

\begin{acknowledgments}
Prof. Dr. Burkhard D\"unweg is gratefully acknowledged for stimulating discussions and critical reading of the manuscript. The author declares no conflict of interest. 
\end{acknowledgments}

\bibliography{ref}

\end{document}